\documentclass[a4paper]{jpconf}
\usepackage{graphicx}
\usepackage{amsmath}
\usepackage{tikz}
\usepackage{placeins}
\usepackage{caption}
\usepackage{color}
\begin{document}

\title{Lattice study of hybrid static potentials}

\author{Philipp Wolf, Marc Wagner}

\address{Goethe-Universit\"at Frankfurt am Main, Institut f\"ur Theoretische Physik, Max-von-Laue-Stra\ss{}e 1, D-60438 Frankfurt am Main, Germany}

\ead{pwolf@th.physik.uni-frankfurt.de}

\begin{abstract}
We report about a recently started project with the aim to compute hybrid static potentials using lattice gauge theory. First preliminary results for pure SU(2) Yang-Mills theory are presented.
\end{abstract}


\section{Introduction}

One of the goals of the PANDA experiment at FAIR will be the search for gluonic excitations, e.g.\ glueballs and hybrid mesons. While glueballs consist only of gluons, hybrid mesons are quark antiquark states with excited gluonic fields, which contribute to the quantum numbers. Hybrid mesons are, therefore, not restricted to quark model quantum numbers: $J^{PC}$ with spin $S=0,1$ and orbital angular momentum $L=0,1,2,...$, where $P=(-1)^{L+1}$ and $C=(-1)^{L+S}$. Thus, exotic states with $J^{PC} = 0^{+-} , 0^{--} , 1^{-+} , \ldots$ can be realized by an excited gluonic field. Examples for $J^{PC} = 1^{-+}$ states are $\pi_1(1400)$ and $\pi_1(1600)$, which are hybrid meson and tetraquark candidates \cite{Yao:2006px}.

The ordinary static potential can be extracted from Wilson loop averages, which are straightforward to compute in lattice gauge theory. Similarly, hybrid static potentials require the computation of Wilson loop-like observables with more complicated spatial transporters, which realize the non-trivial gluonic quantum numbers. For existing lattice studies cf.\ \cite{Juge:1997nc,Peardon:1997jr,Juge:1997ir,
Morningstar:1998xh,Michael:1998tr,Michael:1999ge,Toussaint:1999kh,
Bali:2000vr,Morningstar:2001nu,Juge:2002br,
Michael:2003ai,Juge:2003qd,Michael:2003xg,Bali:2003jq}.


\section{A brief introduction to lattice QCD hadron spectroscopy}

To determine the mass of a hadron, one first has to define a suitable hadron creation operator $\mathcal{O}$ composed of quark and gluon field operators $\psi$, $\bar{\psi}$ and $A_\mu$, which creates the quantum numbers of the hadron of interest, when applied to the vacuum $| \Omega \rangle$. In a second step one computes the temporal correlation function $C(T)$ of this hadron creation operator using lattice QCD,
\begin{eqnarray}
\label{EQN001} C(T) \ \ \equiv \ \ \langle \Omega | \mathcal{O}^\dagger(T) \mathcal{O}(0) | \Omega \rangle \ \ = \ \ \frac{1}{Z} \int \mathcal{D}[A_\mu,\psi,\bar{\psi}] \, \mathcal{O}^\dagger(T) \mathcal{O}(0) e^{-S_{E,\textrm{QCD}}[A_\mu,\psi,\bar{\psi}]} ,
\end{eqnarray}
where $S_{E,\textrm{QCD}}[A_\mu,\psi,\bar{\psi}]$ is the QCD action. The hadron mass $M$ can then be obtained from the exponential decay of $C(T)$ at large temporal separations:
\begin{eqnarray}
C(T) \ \ = \ \ \sum_n \Big|\langle n | \mathcal{O} | \Omega \rangle\Big|^2 e^{-(E_n-E_\Omega) T} \ \ = \ \ \Big|\langle 0 | \mathcal{O} | \Omega \rangle\Big|^2 e^{-M T} \Big(1 + \mathcal{O}(e^{-(E_1-E_0) T)})\Big) ,
\end{eqnarray}
where $| n \rangle$ denote eigenstates of the QCD Hamiltonian with the quantum numbers of the hadron and $M \equiv E_0 - E_\Omega$.

While a straightforward way to determine $M$ is to fit an exponential function $A e^{-M T}$ to the lattice results for $C(T)$ at large temporal separations $T$, it is also common to study the so-called effective mass
\begin{equation}
M_{\textrm{eff}}(T) \ \ \equiv \ \ \frac{1}{a} \ln\bigg(\frac{C(T)}{C(T+a)}\bigg)
\end{equation} 
($a$ denotes the lattice spacing) and to fit a constant to the plateau at sufficiently large $T$,
\begin{equation}
M \ \ =_{\textrm{large }T} \ \ M_{\textrm{eff}}(T) .
\end{equation} 

For a more detailed introduction to lattice hadron spectroscopy cf.\ e.g.\ \cite{Weber:2013eba}.


\section{The static potential}

The static potential $V(R)$ is defined as the energy difference of the lowest state containing a static quark $Q$ and a static antiquark $\bar{Q}$ at separation $r$ and the vacuum. In the absence of quarks of finite mass $V(R)$ is linear for large separations, $\sim \sigma R$, where $\sigma$ denotes the string tension (cf.\ fig.~\ref{FIG002}b), and, therefore, displays the confining property of Yang-Mills theory and QCD.
Quite often it is used to set the scale for lattice simulations, e.g.\ by identifying the resulting lattice string tension with its physical value $\sigma \approx 5.5 / \textrm{fm}^2 \ldots 7.5 / \textrm{fm}^2$.

To compute the static potential, one typically uses the creation operator
\begin{eqnarray}
\label{EQN002} \mathcal{O} \ \ \equiv \ \ \overline{Q}(\mathbf{x}) \mathcal{S}(\mathbf{x},\mathbf{y}) Q(\mathbf{y}) \quad , \quad |\mathbf{x} - \mathbf{y}| \ \ = \ \ R
\end{eqnarray}
where $\overline{Q}(\mathbf{x})$ and $Q(\mathbf{y})$ are operators creating a static antiquark and a static quark at $\mathbf{x}$ and $\mathbf{y}$, respectively. $\mathcal{S}(\mathbf{x},\mathbf{y})$ is a product of spatial gauge links along a straight line connecting $\mathbf{x}$ and $\mathbf{y}$, i.e.\ a parallel transporter realizing a gauge invariant static potential creation operator $\mathcal{O}$.

\begin{figure}[htb]
\begin{minipage}{0.38\textwidth}
\begin{tikzpicture}[scale=.5]
\draw [<->] (0,2.5) -- (0,0) -- (2.5,0);
\node [below] at (2.5,0) {$x$};
\node [left] at (0,2.5) {$t$};
\draw [thick, ->] (1,1) -- (2,1);
\draw [thick] (2,1) -- (3,1);
\draw [thick, ->] (3,1) -- (4,1);
\draw [thick] (4,1) -- (5,1);
\draw [thick, ->] (5,1) -- (5,2);
\draw [thick, ->] (5,3) -- (5,4);
\draw [thick, ->] (5,5) -- (5,6);
\draw [thick] (5,2) -- (5,3);
\draw [thick] (5,4) -- (5,5);
\draw [thick] (5,6) -- (5,7);
\draw [thick, ->] (5,7) -- (4,7);
\draw [thick, ->] (3,7) -- (2,7);
\draw [thick] (4,7) -- (3,7);
\draw [thick] (2,7) -- (1,7);
\draw [thick, ->] (1,7) -- (1,6);
\draw [thick, ->] (1,5) -- (1,4);
\draw [thick, ->] (1,3) -- (1,2);
\draw [thick] (1,6) -- (1,5);
\draw [thick] (1,4) -- (1,3);
\draw [thick] (1,2) -- (1,1);
\shade[shading=ball, ball color=black] (1,1) circle(.1);
\shade[shading=ball, ball color=black] (3,1) circle(.1);
\shade[shading=ball, ball color=black] (5,1) circle(.1);
\shade[shading=ball, ball color=black] (5,3) circle(.1);
\shade[shading=ball, ball color=black] (5,5) circle(.1);
\shade[shading=ball, ball color=black] (5,7) circle(.1);
\shade[shading=ball, ball color=black] (3,7) circle(.1);
\shade[shading=ball, ball color=black] (1,7) circle(.1);
\shade[shading=ball, ball color=black] (1,5) circle(.1);
\shade[shading=ball, ball color=black] (1,3) circle(.1);
\node [left, allow upside down] at (1,4) {$(\mathcal{T}(\mathbf{x};0,t))^\dagger$};
\node [right, allow upside down] at (5,4) {$\mathcal{T}(\mathbf{y};0,t)$};
\node [above] at (3,1) {$\mathcal{S}(\mathbf{x},\mathbf{y};0)$};
\node [above] at (3,7) {$(\mathcal{S}(\mathbf{x},\mathbf{y};t))^\dagger$};
\end{tikzpicture}
\end{minipage}\hspace{1pc}
%
%
\begin{minipage}{0.33\textwidth}
\vspace{-0.32cm}
\begin{tikzpicture}[scale=.5]
\draw [<->] (0,2.5) -- (0,0) -- (2.5,0);
\node [below] at (2.5,0) {$x$};
\node [left] at (0,2.5) {$t$};
\draw [thick] (1,1) -- (2,1);
\draw [ultra thick, red] (2,0.3) rectangle (4,1.8);

\draw [thick] (4,1) -- (5,1);
\draw [thick, ->] (5,1) -- (5,2);
\draw [thick, ->] (5,3) -- (5,4);
\draw [thick, ->] (5,5) -- (5,6);
\draw [thick] (5,2) -- (5,3);
\draw [thick] (5,4) -- (5,5);
\draw [thick] (5,6) -- (5,7);
\draw [thick] (5,7) -- (4,7);
\draw [ultra thick, red] (2,6.3) rectangle (4,7.8);
\node [above right, red] at (4,7.8) {$\dagger$};

\draw [thick] (2,7) -- (1,7);
\draw [thick, ->] (1,7) -- (1,6);
\draw [thick, ->] (1,5) -- (1,4);
\draw [thick, ->] (1,3) -- (1,2);
\draw [thick] (1,6) -- (1,5);
\draw [thick] (1,4) -- (1,3);
\draw [thick] (1,2) -- (1,1);
\shade[shading=ball, ball color=black] (1,1) circle(.1);
\shade[shading=ball, ball color=black] (5,1) circle(.1);
\shade[shading=ball, ball color=black] (5,3) circle(.1);
\shade[shading=ball, ball color=black] (5,5) circle(.1);
\shade[shading=ball, ball color=black] (5,7) circle(.1);
\shade[shading=ball, ball color=black] (1,7) circle(.1);
\shade[shading=ball, ball color=black] (1,5) circle(.1);
\shade[shading=ball, ball color=black] (1,3) circle(.1);

\node at (3,1) {$\mathbf{B, E}$};
\node at (3,7) {$\mathbf{B, E}$};
\end{tikzpicture}
\end{minipage}
%
%
\begin{minipage}{0.25\textwidth}
\vspace{-0.32cm}
\begin{tikzpicture}[scale=.5]
\draw [thick, ->] (0,0) -- (1,0);
\draw [thick] (1,0) -- (2,0);
\draw [thick, ->] (2,0) -- (2,1);
\draw [thick] (2,1) -- (2,2);
\draw [thick, ->] (2,2) -- (1,2);
\draw [thick] (1,2) -- (0,2);
\draw [thick, ->] (0,2) -- (0,1);
\draw [thick] (0,1) -- (0,0);

\draw [thick, ->] (2.3,0) -- (3.3,0);
\draw [thick] (3.3,0) -- (4.3,0);
\draw [thick, ->] (4.3,0) -- (4.3,1);
\draw [thick] (4.3,1) -- (4.3,2);
\draw [thick, ->] (4.3,2) -- (3.3,2);
\draw [thick] (3.3,2) -- (2.3,2);
\draw [thick, ->] (2.3,2) -- (2.3,1);
\draw [thick] (2.3,1) -- (2.3,0);

\draw [thick, ->] (0,2.3) -- (1,2.3);
\draw [thick] (1,2.3) -- (2,2.3);
\draw [thick, ->] (2,2.3) -- (2,3.3);
\draw [thick] (2,3.3) -- (2,4.3);
\draw [thick, ->] (2,4.3) -- (1,4.3);
\draw [thick] (1,4.3) -- (0,4.3);
\draw [thick, ->] (0,4.3) -- (0,3.3);
\draw [thick] (0,3.3) -- (0,2.3);

\draw [thick, ->] (2.3,2.3) -- (3.3,2.3);
\draw [thick] (3.3,2.3) -- (4.3,2.3);
\draw [thick, ->] (4.3,2.3) -- (4.3,3.3);
\draw [thick] (4.3,3.3) -- (4.3,4.3);
\draw [thick, ->] (4.3,4.3) -- (3.3,4.3);
\draw [thick] (3.3,4.3) -- (2.3,4.3);
\draw [thick, ->] (2.3,4.3) -- (2.3,3.3);
\draw [thick] (2.3,3.3) -- (2.3,2.3);

\shade[shading=ball, ball color=black] (2.15,2.15) circle(.25);
\end{tikzpicture}
\end{minipage}
\caption{\label{FIG001}
\textbf{(a)}~Ordinary Wilson loop $W(R=2,T=3)$ in the $x$-$t$ plane.
\textbf{(b)}~Generalized Wilson loop: $\mathbf{B}$ or $\mathbf{E}$ fields inserted at the centers of the spatial transporters.
\textbf{(c)}~Average of neighboring plaquettes representing a chromomagnetic or chromoelectric field $\mathbf{B}$ or $\mathbf{E}$.
}
\end{figure}
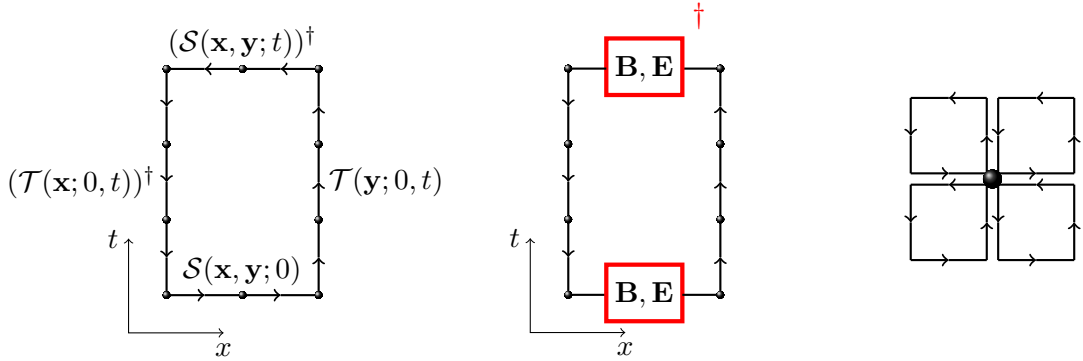

\begin{figure}[htb]
\begin{center}
\includegraphics[width=0.35\textwidth]{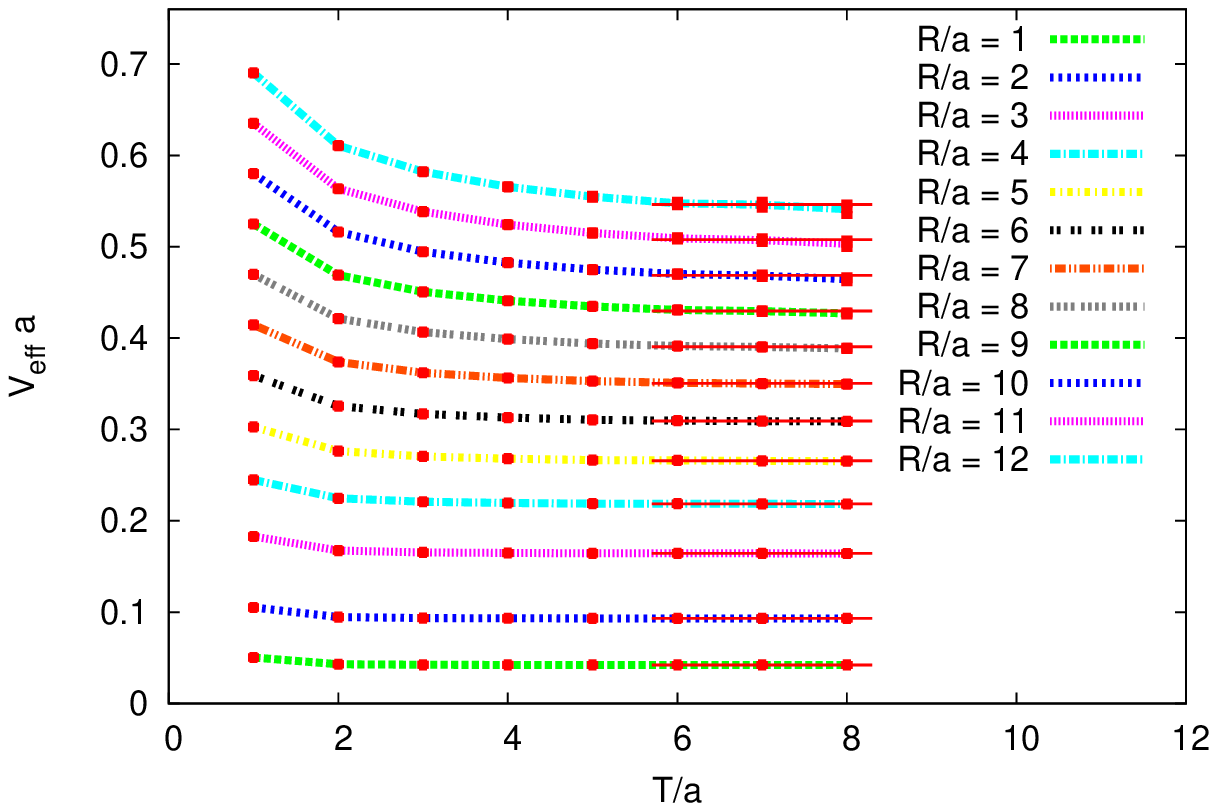}
\includegraphics[width=0.35\textwidth]{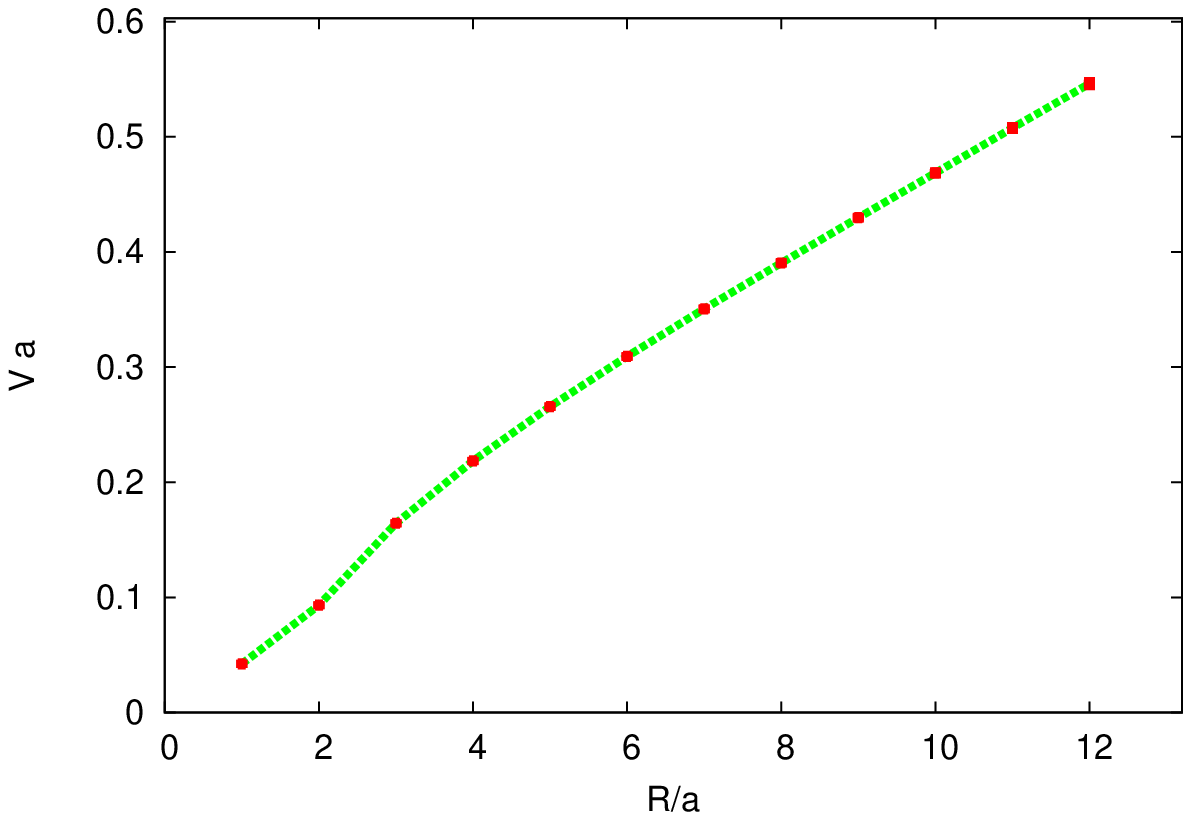}
\end{center}
\caption{\label{FIG002}
\textbf{(a)}~Effective potential $V_{\textrm{eff}}(R,T)$ as a function of $T$ for different quark antiquark separations $R$ in units of the lattice spacing $a \approx 0.073 \, \textrm{fm}$.
\textbf{(b)}~Static potential $V(R)$ in units of the lattice spacing.
}
\end{figure}

The integration over the static quark fields $Q$ and $\bar{Q}$ in the path integral (\ref{EQN001}) can be performed analytically. In a computation without dynamical sea quarks the result is
\begin{eqnarray}
C(T) \ \ \propto \ \ \Big\langle W(R,T) \Big\rangle \ \ \equiv \ \ \frac{1}{Z} \int \mathcal{D}[A_\mu] \, W(R,T) e^{-S_{E,\textrm{YM}}[A_\mu]}
\end{eqnarray}
with the Yang-Mills action $S_{E,\textrm{YM}}[A_\mu]$. $W(R,T)$ is a product of gauge links along a closed rectangular path of spatial extent $R$ and temporal extent $T$, the well known Wilson loop. For a precise definition in terms of gauge links it can be decomposed in four parts, two spatial parallel transporters $\mathcal{S}$ and two temporal parallel transporters $\mathcal{T}$,
\begin{equation*}
\mathcal{S}(\mathbf{x},\mathbf{y};t) \equiv \prod_{j=0}^{R/a-1} U_k(\mathbf{x}+j a \mathbf{e}_k,t) \quad , \quad \mathcal{T}(\mathbf{x};t_1,t_2) \ \ \equiv \ \ \prod_{j=0}^{T/a-1} U_0(\mathbf{x},t_1+j a) .
\end{equation*}
Then
\begin{eqnarray}
W(R,T) \ \ \equiv \ \ \text{Tr}\Big(
\mathcal{S}(\mathbf{x},\mathbf{y};0)
\mathcal{T}(\mathbf{y};0,t)
\Big(\mathcal{S}(\mathbf{x},\mathbf{y};t)\Big)^\dagger
\Big(\mathcal{T}(\mathbf{x};0,t)\Big)^\dagger
\Big)
\end{eqnarray}
(cf.\ fig.~\ref{FIG001}a).

For each quark antiquark separation $R$ an effective mass or effective potential
\begin{equation}
V_{\textrm{eff}}(R,T) \ \ \equiv \ \ \frac{1}{a} \ln\bigg(\frac{\langle W(R,T) \rangle}{\langle W(R,T+a) \rangle}\bigg)
\end{equation} 
has to be computed (cf.\ fig.~\ref{FIG002}a) as explained in the previous section. Fitting a constant to $V_{\textrm{eff}}(R,T)$ in the plateau-like region at $T/a = 6 \ldots 8$ yields the static potential $V(R)$ (cf.\ fig.~\ref{FIG002}b).

\section{Hybrid static potentials}

States with a static quark and a static antiquark can be classified according to the following three quantum numbers (for a more detailed discussion cf.\ e.g.\ \cite{Bali:2005fu,Wagner:2010ad}):
\begin{itemize}
\item Angular momentum $J$ with respect to the axis of separation of the quark antiquark pair. States with $J=0,\pm 1, \pm 2, \ldots$ are also labeled by $\Sigma, \Pi, \Delta, \ldots$.

\item The combination of parity and charge conjugation $P \circ C$. States with $P \circ C = +,-$ are also labeled by $g,u$.

\item The rotational invariant $\Sigma$ states are either symmetric or antisymmetric with respect to spatial reflections along an axis perpendicular to the axis of separation of the quark antiquark pair denoted by $P_x = +,-$.
\end{itemize}
For example the ordinary static potential discussed in the previous section has quantum numbers $J_{P \circ C}^{P_x} = \Sigma_g^+$.

Hybrid static potentials correspond to a quark antiquark configuration with an excited gluonic string realizing quantum numbers different from $\Sigma_g^+$. One way to implement such quantum numbers in a hybrid static potential creation operator similar to (\ref{EQN002}) is to insert a suitable gluonic operator at the center of the spatial parallel transporter $\mathcal{S}$. Typically this insertion is a chromomagnetic or chromoelectric field operator $\mathbf{B}$ or $\mathbf{E}$ possibly combined with the separation vector $\mathbf{R} \equiv \mathbf{y}-\mathbf{x}$ and/or the covariant derivative $\mathbf{D}$ (cf.\ fig.~\ref{FIG001}b). On a lattice $\mathbf{B}$ and $\mathbf{E}$ are expressed in terms of the average of four neighboring plaquettes of gauge links (cf.\ fig.~\ref{FIG001}c). The creation operators used in this work are collected in tab.~\ref{TAB001}. A more detailed discussion of such operators can e.g.\ be found in \cite{Juge:2002br}.

\begin{table}[htb]
\scriptsize
\begin{center}
\begin{tabular}{cc}
\br 
quantum numbers $J_{P \circ C}^{P_x}$ & operator insertions \\
\mr
$\Sigma_g^+$ & $1 \quad , \quad \mathbf{R} \cdot \mathbf{E} \quad , \quad \mathbf{R} \cdot (\mathbf{D} \times \mathbf{B})$ \\
$\Pi_g$ & $\mathbf{R} \times \mathbf{E} \quad , \quad \mathbf{R} \times (\mathbf{D} \times \mathbf{B})$ \\
\mr
$\Sigma_u^-$ & $\mathbf{R} \cdot \mathbf{B} \quad , \quad \mathbf{R} \cdot (\mathbf{D} \times \mathbf{E})$ \\
$\Pi_u$ & $\mathbf{R} \times \mathbf{B} \quad , \quad \mathbf{R} \times (\mathbf{D} \times \mathbf{E})$ \\
\mr
$\Sigma_g^-$ & $(\mathbf{R} \cdot \mathbf{D})(\mathbf{R} \cdot \mathbf{B})$ \\
\br 
\end{tabular}
\end{center}
\caption{\label{TAB001}Hybrid static potential creation operators.}
\end{table}

\vspace{-0.9cm}

\section{Preliminary numerical results}


\subsection{Lattice setup}

We have evaluated Wilson loops and generalized Wilson loops on more than 700 essentially independent gauge link configurations with $24^4$ lattice sites. These configurations have been generated with the SU(2) standard Wilson plaquette action using a heatbath algorithm and gauge coupling $\beta = 2.50$. The corresponding lattice spacing is $a \approx 0.073 \, \textrm{fm}$, when identifying the Sommer parameter $r_0$ with $0.46 \, \textrm{fm}$ \cite{Philipsen:2013ysa}.


\subsection{Hybrid static potentials}

\begin{figure}[htb]
\begin{center}
\includegraphics[width=0.58\textwidth]{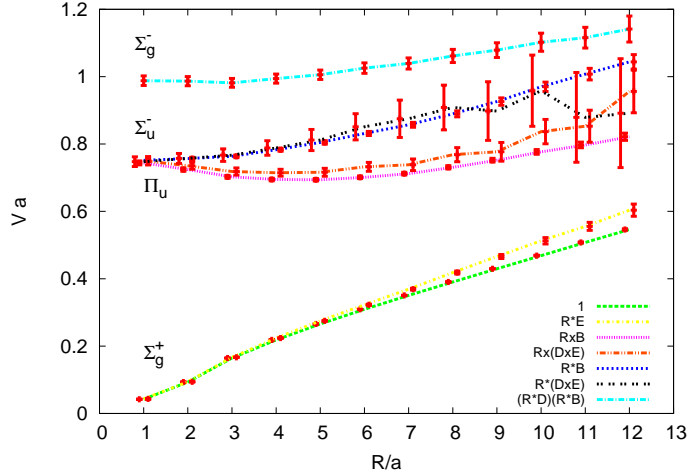}
\end{center}
\caption{\label{FIG003}Hybrid static potentials in SU(2) Yang-Mills theory in units of the lattice spacing $a \approx 0.073 \, \textrm{fm}$ (to be able to distinguish different curves, error bars have slightly been shifted horizontally).}
\end{figure}

In fig.~\ref{FIG003} our preliminary results for hybrid static potentials with quantum numbers $\Sigma_g^+$, $\Pi_u$, $\Sigma_u^-$ and $\Sigma_g^-$ are shown. For each of the sectors $\Sigma_g^+$, $\Pi_u$ and $\Sigma_u^-$ two different creation operators have been used resulting, as expected, in potentials, which are identical within statistical errors. For example the ordinary static potential with $J_{P \circ C}^{P_x} = \Sigma_g^+$ has been extracted from Wilson loops (green curve) and from generalized Wilson loops with insertions $\mathbf{R} \cdot \mathbf{E}$ (yellow curve). Clearly the green curve is of better quality, i.e.\ exhibits less fluctuations and smaller statistical errors. Consequently, ordinary Wilson loops are much better suited to determine the $\Sigma_g^+$ static potential than Wilson loops with insertions $\mathbf{R} \cdot \mathbf{E}$. Similarly, $\mathbf{R} \times \mathbf{B}$ (magenta curve) is superior to $\mathbf{R} \times (\mathbf{D} \times \mathbf{E})$ (orange curve), when computing the $\Pi_u$ static potential, and $\mathbf{R} \cdot \mathbf{B}$ (blue curve) is superior to $\mathbf{R} \cdot (\mathbf{D} \times \mathbf{E})$ (black curve), when computing the $\Sigma_u^-$ static potential. For small separations the potentials $\Pi_u$ and $\Sigma_g^-$ seem to be repulsive. To exclude that this is merely a cutoff effect, which are expected to be large for $R/a < 2$, we plan to perform similar computations at significantly smaller lattice spacing in the near future.

\section{Outlook}

Even though we have used a sizable number of more than 700 gauge link configurations, the statistical errors of our hybrid static potential results shown in fig.~\ref{FIG003} are quite large, when e.g.\ compared to results from \cite{Juge:2002br}. This might be due to a possibly different structure of the employed hybrid static potential creation operators. While for our current results we have used local insertions of $\mathbf{B}$ and $\mathbf{E}$ field operators, which might generate a rather poor ground state overlap, in \cite{Michael:1998tr,Michael:1999ge} spatially extended creation operators have been proposed, which could result in correlation functions, which are dominated by the ground state already at small temporal separations. We plan to explore this issue in the near future by implementing additional hybrid static potential creation operators.

The ultimate goal is, of course, to arrive at precise results for SU(3) Yang-Mills theory and QCD. This would allow to estimate masses of hybrid mesons by solving a non-relativistic Schr\"odinger equation with the computed hybrid static potentials, as e.g.\ proposed in \cite{Peardon:1997jr,Juge:1997ir,Michael:1998tr,Michael:1999ge,
Michael:2003ai,Juge:2003qd,Michael:2003xg}. Moreover, in the context of effective field theories like pNRQCD there is considerable interest in the short distance behavior of hybrid static potentials, which is related to gluelump masses \cite{Bali:2003jq,Brambilla:2004jw}. It would also be interesting to compare such lattice results with corresponding model calculations, e.g.\ \cite{Andreev:2012mc}.


\ack

We acknowledge helpful discussions with Joshua Berlin, Owe Philipsen and Bj\"orn Wagenbach. M.W.\ acknowledges support by the Emmy Noether Programme of the DFG (German Research Foundation), grant WA 3000/1-1. This work was supported in part by the Helmholtz International Center for FAIR within the framework of the LOEWE program launched by the State of Hesse.




\section*{References}

\begin{thebibliography}{99}

\bibitem{Yao:2006px} 
  W.~M.~Yao {\it et al.}  [Particle Data Group Collaboration],
  J.\ Phys.\ G {\bf 33}, 1 (2006).

\bibitem{Juge:1997nc} 
  K.~J.~Juge, J.~Kuti and C.~J.~Morningstar,
  Nucl.\ Phys.\ Proc.\ Suppl.\ {\bf 63}, 326 (1998)
  [hep-lat/9709131].

 \bibitem{Peardon:1997jr} 
  M.~J.~Peardon,
  Nucl.\ Phys.\ Proc.\ Suppl.\  {\bf 63}, 22 (1998)
  [hep-lat/9710029].

 \bibitem{Juge:1997ir} 
  K.~J.~Juge, J.~Kuti and C.~J.~Morningstar,
  AIP Conf.\ Proc.\  {\bf 432}, 136 (1998)
  [hep-ph/9711451].
  
  
 \bibitem{Morningstar:1998xh} 
  C.~Morningstar, K.~J.~Juge and J.~Kuti,
  hep-lat/9809015.

\bibitem{Michael:1998tr} 
  C.~Michael,
  Nucl.\ Phys.\ A {\bf 655}, 12 (1999)
  [hep-ph/9810415].

\bibitem{Toussaint:1999kh} 
  D.~Toussaint,
  Nucl.\ Phys.\ Proc.\ Suppl.\  {\bf 83}, 151 (2000)
  [hep-lat/9909088].

\bibitem{Michael:1999ge} 
  C.~Michael,
  PoS {\bf HeavyFlavours8}, 001 (1999)
  [hep-ph/9911219].

\bibitem{Bali:2000vr} 
  G.~S.~Bali {\it et al.} [SESAM and T$\chi$L Collaborations],
  Phys.\ Rev.\ D {\bf 62}, 054503 (2000)
  [hep-lat/0003012].

\bibitem{Morningstar:2001nu} 
  C.~Morningstar,
  AIP Conf.\ Proc.\ {\bf 619}, 231 (2002)
  [nucl-th/0110074].

\bibitem{Juge:2002br} 
  K.~J.~Juge, J.~Kuti and C.~Morningstar,
  Phys.\ Rev.\ Lett.\  {\bf 90}, 161601 (2003)
  [hep-lat/0207004].

  \bibitem{Michael:2003ai} 
  C.~Michael,
  hep-lat/0302001.

\bibitem{Juge:2003qd} 
  K.~J.~Juge, J.~Kuti and C.~Morningstar,
  AIP Conf.\ Proc.\ {\bf 688}, 193 (2004)
  [nucl-th/0307116].

  \bibitem{Michael:2003xg} 
  C.~Michael,
  hep-ph/0308293.
  

\bibitem{Bali:2003jq} 
  G.~S.~Bali and A.~Pineda,
  Phys.\ Rev.\ D {\bf 69}, 094001 (2004)
  [hep-ph/0310130].

\bibitem{Weber:2013eba} 
  M.~Wagner, S.~Diehl, T.~Kuske and J.~Weber,
  arXiv:1310.1760 [hep-lat].

\bibitem{Bali:2005fu} 
  G.~S.~Bali {\it et al.} [SESAM Collaboration],
  Phys.\ Rev.\ D {\bf 71}, 114513 (2005)
  [hep-lat/0505012].

\bibitem{Wagner:2010ad} 
  M.~Wagner [ETMC Collaboration],
  PoS LATTICE {\bf 2010}, 162 (2010)
  [arXiv:1008.1538 [hep-lat]].

\bibitem{Philipsen:2013ysa} 
  O.~Philipsen and M.~Wagner,
  Phys.\ Rev.\ D {\bf 89}, 014509 (2014)
  [arXiv:1305.5957 [hep-lat]].

\bibitem{Brambilla:2004jw} 
  N.~Brambilla, A.~Pineda, J.~Soto and A.~Vairo,
  Rev.\ Mod.\ Phys.\ {\bf 77}, 1423 (2005)
  [hep-ph/0410047].
  
\bibitem{Andreev:2012mc} 
  O.~Andreev,
  Phys.\ Rev.\ D {\bf 86}, 065013 (2012)
  [arXiv:1207.1892 [hep-ph]].
 
\end{thebibliography}
\end{document}